\documentclass{aa}
\usepackage{epsf,epsfig}
\begin{document}

\title{A metallicity-flattening relation for dwarf elliptical galaxies}

\author{Fabio D. Barazza \inst{}
\and Bruno Binggeli \inst{}}

\offprints{F.D. Barazza, e-mail : barazza@astro.unibas.ch}   

\institute{Astronomisches Institut, Universit\"at Basel, Venusstrasse 7,
CH-4102 Binningen, Switzerland}

\date{Received / Accepted}

\abstract{It is shown that bright cluster dwarf ellipticals follow a 
relation between metallicity and apparent flattening. Rounder dwarfs tend to 
be more metal-rich. The evidence is based on colour as well as spectroscopic 
line-strength data from the literature. The large scatter of dEs
around the mean metallicity-luminosity relation, usually ascribed to
large observational errors, turns out to be an ellipticity effect.
In the magnitude range $-14 \ga M_B \ga -18$
the metallicity of dEs depends more strongly on ellipticity than luminosity.
A possible explanation is that galaxies with masses around 
$10^{9}$ $M_{\odot}$ suffered a partial blowout of metal-enriched gas along
their minor axis, rendering ellipticity a critical parameter for metallicity
(De Young \& Heckman 1994). 
}
\maketitle 

\section{Introduction}
Elliptical galaxies follow a well-known luminosity-metallicity
relation, in the sense that more luminous ellipticals are observed to be
globally redder, or 
have larger metallic absorption line strengths in their spectra than less 
luminous systems (Faber 1973, Visvanathan \& Sandage 1977). Obviously this is  
a mass-metallicity relation, and its canonical interpretation is that less
massive galaxies had more significant outflows 
of metal-enriched gas at an 
early evolutionary stage (Faber 1973, Mould 1984). The relation seems to hold
for the entire luminosity range of spheroidal galaxies, down to the 
faintest dwarfs (Caldwell 1983, Brodie \& Huchra 1991, Caldwell et 
al.~1992). However, there is also considerable scatter in the relation, 
and there have been attempts to find a {\em second parameter}\/ which 
governs metallicity (mass being the primary parameter). A good candidate
is the internal velocity dispersion (Terlevich et al.~1984, 
Efstathiou \& Fall 1984), which even more strongly correlates 
with metallicity than mass (Bender et al.~1993). 

The largest scatter in the luminosity-metallicity relation is observed in the
magnitude range of bright cluster dwarf ellipticals ($-14 \ga M_B \ga -18$).
While this scatter could plausibly be due to measurement errors,
Rakos et al.~(2001) find a weak correlation with age for a sample of Fornax
cluster dwarfs observed in Stroemgren narrow-band colours.
 
In this letter we show that the scatter in metallicity 
for cluster dEs is largely explained by {\em apparent 
ellipticity}\/ (flattening). At a given luminosity, {\em rounder
dwarf ellipticals are more metal-rich}.
There have already been hints that ellipticity
might act as a second parameter in normal elliptical galaxies (Terlevich 
et al.~1984). However, for dEs the effect is so strong that ellipticity 
appears to be the {\em primary}\/ parameter. 
Such a metallicity-flattening relation for dEs is not implausible.
The outflow of metal-enriched gas in stellar systems of intermediate mass
($M \approx 10^{9}$ $M_{\odot}$) is 
preferentially occurring along the minor axis,
rendering ellipticity a critical parameter for the metallicity of present
day dwarfs (De Young \& Heckman 1994). {\em Rounder dEs seem to
have suffered less significant outflow}.

\section{Colour versus flattening}
We first show the evidence from multi-colour photometry, 
which is how we stumbled over the flattening effect.
In Fig.~1 we have plotted integrated colour
versus apparent ellipticity for different samples of spheroidal galaxies. 
The filled circles are $U-B$ values for 15 Virgo dEs
from our own photometry (Barazza et al., in
prep.). There is a very clear trend appearing, in the sense that 
rounder dEs are redder than more flattened
ones. Such a correlation is rather unexpected -- see for comparison
a sample of normal ellipticals from Peletier et al. (1990, 
triangles). However, our finding is corroborated by the $U-B$ data for
Virgo and Fornax dEs (crossed circles) from Caldwell (1983) and Caldwell \& 
Bothun (1987), and the Stroemgren $vz-yz$ data for Fornax dEs (open circles)
from Rakos et al.~(2001). In contrast, local dwarfs with colour data available
(from Mateo 1998, asterisks) show no clear relation. All ellipticities used
here are taken either from the references given for colour/metallicity data,
or from LEDA.

Is the trend for cluster dEs real? 
Observational bias can be excluded: the samples were not
selected by colour or ellipticity. Nor is it conceivable that the photometric
errors in colour somehow depend on the ellipticity of the whole galaxy.
Another suspicion is that ellipticity is not an independent 
parameter: if ellipticity were correlated
with absolute magnitude, we simply might have recovered the well-known
luminosity-colour relation (e.g.~Ferguson 1994). 
However, the ellipticity of dEs is not significantly related to
luminosity, nor to surface brightness (Binggeli \& Popescu 1995). 

We take it for granted that colour is essentially indicating metallicity, 
as spheroidal galaxies are believed to be dust-free and old.
Of course, metallicity can also be measured more directly.
We now show that the 
colour-ellipticity relation is indeed a metallicity-ellipticity relation.
\begin{figure}[t]
\begin{center}
\epsfig{file=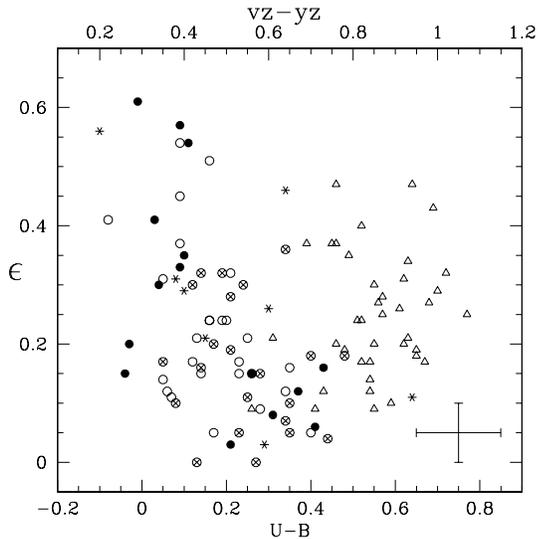,height=75mm,width=75mm}
\end{center}
\caption[]{Colour versus apparent ellipticity for cluster dEs (circles), 
local dEs/dSphs (asterisks), and normal ellipticals (triangles).
Filled circles are Virgo dEs from our own photometry, crossed circles are 
Virgo and Fornax dEs from Caldwell et al., open circles are Fornax dEs
from Rakos et al.~(complete references given in the text). For the Rakos et
al.~data the colour plotted is $vz-yz$ (Stroemgren system), shifted by an
arbitrary amount along the abscissa (scale on top).
For all other samples the conventional $U-B$ colour is plotted (bottom 
scale). Typical error bars are shown in the lower right.}
\end{figure}
\begin{figure}[t]
\begin{center}
\epsfig{file=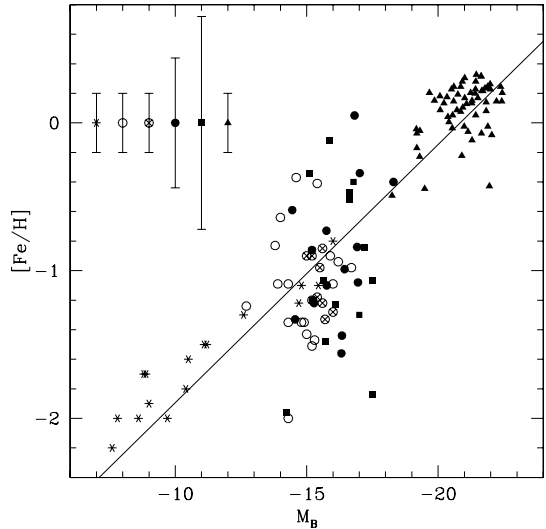,height=75mm,width=75mm}
\end{center}
\caption[]{[Fe/H] versus absolute blue magnitude for
cluster dEs (circles and squares), local dEs/dSphs (asterisks), and
normal Es (small triangles). Open circles are estimated [Fe/H] values 
for Fornax dEs based on Stroemgren colours (Rakos et al.), crossed circles 
are spectroscopically measured metallicities for Fornax dEs (Held \& Mould).
Filled circles and squares are [Fe/H] values for Virgo and Fornax dEs, 
spectroscopically determined as well (Brodie \& Huchra), 
but with considerable errors.
Mean errors for the different samples are shown as error bars in the
upper left corner. Complete references are given in the text. The line is
a fit to all data (equation given in text).}
\end{figure}

\section{The luminosity-metallicity relation}
Fig.~2 shows the ``universal'' luminosity-metallicity relation
for essentially the same galaxies as in Fig.~1. 
Normal Es (small triangles) are represented by data from a more
recent study (Kobayashi \& Arimoto 1999). [Fe/H] values for local dwarfs
(asterisks) are again from Mateo (1998). For bright cluster dEs
($-14 \ga M_B \ga -18$) we have three sources for [Fe/H]:
1) Brodie \& Huchra (1991) give spectroscopically determined metallicities
for nearly 40 Virgo and Fornax cluster dEs (essentially the 
Caldwell et al.~sample) based on several different line 
indices. However, the estimated errors in [Fe/H] are considerable. 
Data with 1\,$\sigma$ errors less
than 0.6 dex are plotted as filled circles, those with larger errors (but
smaller than 1 dex) as filled squares. 2) Held \& Mould (1994)
provide spectroscopically measured [Fe/H] values, with much smaller 
estimated errors, for 8 nucleated Fornax dEs (crossed circles).
3) Finally, there are the Rakos et al.~(2001) $vz-yz$ Stroemgren colours for
Fornax dEs, appropriately transformed into [Fe/H] values (open circles), 
with a small mean error of 0.2 dex.

Naively fitting a line through all 
data plotted in Fig.~2, we get an expression
for the mean (universal) luminosity-metallicity relation:
[Fe/H] = $-0.1747 \, M_B - 3.6405$, which will be used   
for a residual analysis below. The local dwarfs follow the mean relation
surprisingly well. In contrast, the cluster dEs in the intermediate
range between normal Es and extreme dwarfs, while falling in place with the
universal relation in the mean, show enormous scatter in their
{\em individual}\/ metallicities.
The natural suspicion is that this is simply due to large errors in 
[Fe/H]. However, we now show that this scatter
is systematically related to ellipticity, i.e.~it must, at least partially,
be real.

\section{Flattening as second parameter}
The residual metallicities with respect to the line shown in Fig.~2
are plotted versus ellipticity in Fig.~3.
There is clearly no correlation for normal Es and local dwarfs
(upper panel), whereas a relation is evident for cluster dEs (lower
panel): dEs with positive residuals are throughout round,
while highly flattened dEs have always negative residuals.
At a given luminosity, rounder dEs tend to be more metal-rich. 
\begin{figure}[t]
\begin{center}
\epsfig{file=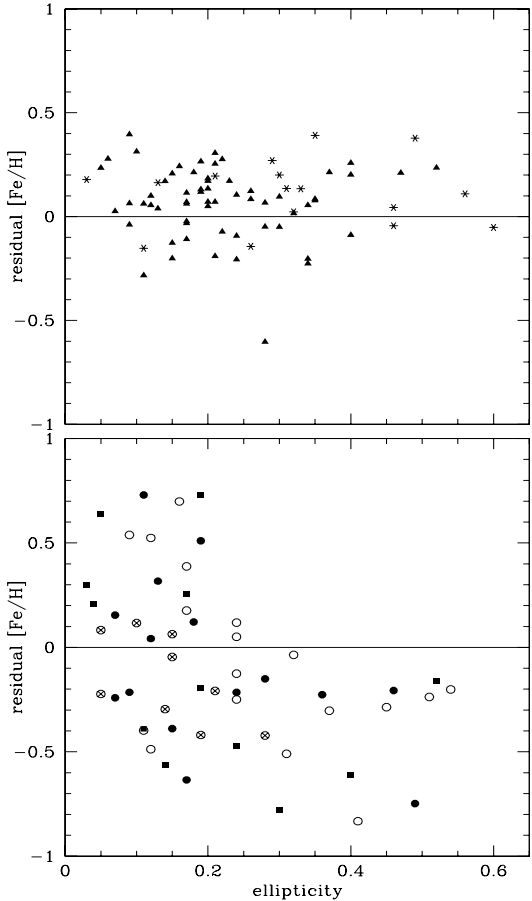,height=120mm,width=70mm}
\end{center}
\caption[]{Residual [Fe/H] with respect to the linear fit in Fig.~2 versus
apparent ellipticity for 
normal ellipticals and local dwarfs (upper panel), and for cluster dEs
(lower panel). Symbols as in Fig.~2.}
\end{figure}

Several points have to be noted. First,  
the trend is followed by {\em all}\/ dE samples used, 
even those where large errors
were claimed (the Brodie \& Huchra data). Again it is not conceivable
why these errors, were they real, should correlate with ellipticity
(we believe the errors are simply overestimated).
Second, we have tested that the metallicity residuals do not correlate with
other parameters, such as
effective radius, i.e. there is no other ``second'' parameter
than ellipticity. Third, the true physical relation behind this 
effect will involve {\em intrinsic}\/ ellipticity. Due to random projection
this relation should even be {\em stronger}\/ than what we see with 
apparent ellipticity. The distribution of points in Fig.~3 (lower panel) is
in accord with what we expect from
projection effects: apparently round galaxies with positive residuals 
will also be intrinsically round, while some of the 
apparently round galaxies with negative residuals will intrinsically be more
flattened (i.e.~would be shifted to the right). 

\section{The metallicity-flattening relation}
\begin{figure}[t]
\begin{center}
\epsfig{file=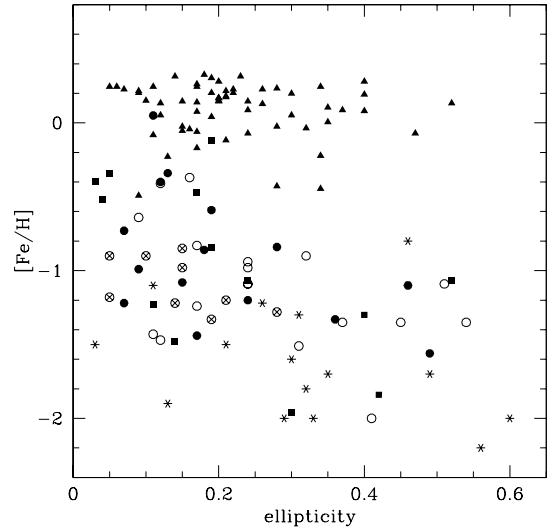,height=75mm,width=75mm}
\end{center}
\caption[]{[Fe/H] versus apparent ellipticity. Symbols as in Fig.~2.}
\end{figure}
In Fig.~4 we have plotted metallicity directly versus ellipticity, 
to get an equivalent to the colour-ellipticity diagram of Fig.~1. As expected,
no correlation is seen for giant Es, while dwarf objects evidently follow
the suspected relation: rounder galaxies tend to be more metal-rich than 
flatter ones. Owing to some outliers, the trend is less evident for
the local dwarfs (no correlation at all was seen in the residual plot). 
Conceivably, local ({\em field}) dwarfs are systematically different from 
{\em cluster}\/ dwarfs (see Sect.~6).  

In general, the flattening-metallicity relation is not as striking as 
the residual plots in Fig.~3. This is because
luminosity now acts as a hidden parameter, adding scatter. In Fig.~5
we have plotted metallicity versus logarithmic ellipticity for 
only cluster dwarfs, which provides an almost linear relation. A linear 
fit yields
[Fe/H]=$-0.72\log(\epsilon) -1.56$, or in terms of abundance:
$Z/Z_{\odot} \propto \epsilon^{-0.72}$.
\begin{figure}[t]
\begin{center}
\epsfig{file=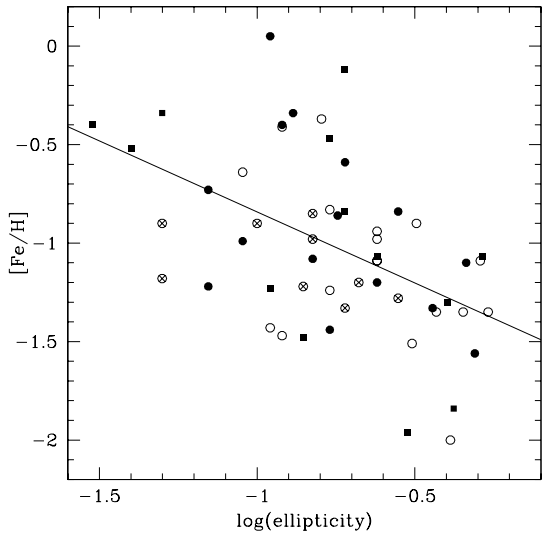,height=75mm,width=75mm}
\end{center}
\caption[]{[Fe/H] versus log ellipticity for cluster dEs.
The line is a linear fit to the data (equation given in text). Symbols as in
Fig.~2.}
\end{figure}
Including the absolute magnitude and fitting a plane to the cluster dwarfs in
the parameter space defined by [Fe/H], $\epsilon$ and $M_B$, we get
$Z/Z_{\odot} \propto \epsilon^{-0.70} L^{-0.20}$, which clearly shows that 
ellipticity is in fact the {\em primary}\/ parameter, 
having a stronger effect on the metallicity of dEs than the luminosity.

\section{Discussion of a possible explanation: outflow of metal-rich gas along
the minor axis}
The mass-metallicity relation of spheroidal galaxies, evident in Fig.~2, is 
most likely the result of early chemodynamical evolution. 
The central idea is gas
loss by a galactic wind, which is believed to be particularly metal-enhanced 
(Vader 1986, 1987). Certainly, the mass (i.e.~potential depth) of a galaxy
must be a key parameter for the loss of gas (hence metals), and this 
explains the mass-metallicity relation. However, it has been pointed out by
De Young \& Heckman (1994) that the shape (i.e.~flattening) of a galaxy 
may play a crucial role as well. These authors   
show that galaxies of intermediate mass ($M \approx 10^9 M_{\odot}$)
would not lose almost all of their ISM in a 
central starburst, but would preferentially
undergo a blowout event, having gas flowing out in the direction of 
their minor axis. The important point is that the 
strength of such an outflow would critically depend on the intrinsic flattening
of the ISM distribution, i.e.~the galaxy. Our
cluster dwarfs showing a metallicity-flattening relation have
masses precisely in the critical regime discussed by De Young \& Heckman
(1994): their mass seems to be too low to be completely protected
from gas loss (as in giant ellipticals), but also too high to experience a 
total blowaway (as in dwarf spheroidals). In the intermediate mass range
of bright cluster dEs the ellipticity could conceivably 
be the dominant parameter for gas loss, and hence metallicity.

Of course, reality must be more complicated. For one thing dwarf galaxies are 
believed to have large amounts of dark
matter, so the gas escaping the stellar body should 
remain bound to the galaxy and be re-accreted on a rather short time scale
(Mac Low \& Ferrara 1999, Ferrara \& Tolstoy 2000). In this context, it
is interesting to note that we have found a possible difference between cluster
dwarfs and (local) field dwarfs. While the evidence for dark matter in field
dwarfs abounds, the situation with cluster galaxies in general is much less
clear. X-ray and lensing studies of clusters of galaxies suggest that 
the dark matter in clusters is not bound to individual galaxies, hence cluster
dwarfs might be much less dark than field dwarfs. An additional difference   
is that cluster dwarfs are subject to ram-pressure stripping by the ICM, 
which might help to remove the gas flown out (e.g.~Murakami \& Babul (1999).

\section{Conclusion}
We have presented evidence for a metallicity-flattening relation for dwarf
elliptical galaxies based on colour and metallicity data available from the
literature. At a given total magnitude, rounder dEs are more metal-rich.
In the narrow magnitude range of bright cluster dwarfs ($-14 \ga M_B \ga -18$),
metallicity is more strongly correlated with ellipticity than luminosity
(mass), i.e. ellipticity seems to be the primary parameter for the 
enrichment history of these galaxies. Possibly this holds only
for {\em cluster}\/ dwarfs; the effect is not significant for local dwarfs.
A possible explanation is provided by the scenario of De Young \&
Heckman (1994), where the outflow of gas, and hence the regulation of 
metallicity, depends on the intrinsic shape of a galaxy in the intermediate
mass range around $10^9 M_{\odot}$.

It would be highly desirable to strengthen the evidence with further 
observations, especially spectroscopically well determined metallicities
for many cluster dwarfs. If confirmed, the effect is likely to be of importance
for our understanding of the chemodynamical evolution of galaxies.

\begin{acknowledgements}
We are grateful to the Swiss National Science Foundation for
financial support. We have made use of the LEDA database
(http://leda.univ-lyon1.fr).  
\end{acknowledgements}

\end{document}